\documentclass[preprint,12pt]{elsarticle}

\setcitestyle{numbers} 

\usepackage{amssymb}
\usepackage{amsmath}

\usepackage{subcaption}

\usepackage{adjustbox}

\journal{Radiation Physics and Chemistry}

\begin{document}

\begin{frontmatter}

\title{\textbf{Ambient neutron dosimetry in high energy and pulsed neutron fields}} 


\author[label1,label2]{Ariel Tarifeno-Saldivia} 
\author[label2]{Francisco Calvino}
\author[label2]{Alfredo  De Blas}
\author[label2]{Benedetta Brusasco}
\author[label2]{Adrià Casanovas-Hoste}
\author[label3]{Ana M. Cives}
\author[label2]{Guillem	Cortés}
\author[label2]{Roger 	García}
\author[label4]{Francisco	Molina}
\author[label2]{Nil 	Mont-Geli}
\author[label2]{Max 	Pall\`as}
\author[label5]{Cristian 	Pavez}

\affiliation[label1]{organization={Instituto de Física Corpuscular (CSIC-UV)},
            addressline={Catedrático José Beltrán 2}, 
            city={Paterna},
            postcode={E-46980}, 
            country={España}}

\affiliation[label2]{
organization={Institut de Tècniques Energètiques, Universitat Politècnica de Catalunya},
            addressline={Av. Diagonal 647}, 
            city={Barcelona},
            postcode={E-08028}, 
            country={España}}

\affiliation[label3]{organization={Centro de Láseres Pulsados (CLPU)},
             addressline={Edicio M5. Parque Cientco. C/ Adaja, 8},
             city={Villamayor, Salamanca},
             postcode={37185},
             country={Spain}}

\affiliation[label4]{organization={Centro de Investigación en Física Nuclear y Espectroscopia de Neutrones CEFNEN. Comisión Chilena de Energía Nuclear.},
             addressline={Nueva Bilbao 12501},
             city={Las Condes, Santiago Chile},
             country={Chile}}

\affiliation[label5]{organization={Comisión Chilena de Energía Nuclear, Center for Research in the Intersection of Plasma Physics, Matter and Complexity, P2mc},
             addressline={Nueva Bilbao 12501},
             city={Las Condes, Santiago Chile},
             country={Chile}}

\begin{abstract}
The status of the LINrem project is presented, focusing on the development of innovative neutron dosimeters with enhanced energy sensitivity, time resolution, and portability. Designed to meet the technical demands of radiation protection in modern particle and nuclear facilities, these dosimeters are discussed in detail. Results from experimental campaigns showcasing their efficacy in pulsed fields generated by fusion plasmas and high-intensity pulsed lasers are presented. Additionally, prospects and future plans for the LINrem project are outlined. 
\end{abstract}

\begin{keyword}
Neutron dosimeters \sep detector design \sep pulsed fields \sep high energy neutron fields


\end{keyword}

\end{frontmatter}


\section{Introduction}
Monitoring neutron dose rates is crucial to ensure minimal risks for workers, patients, and the public in different facilities. At the level of operational radiological protection, the most common monitors are known as ambient neutron dosimeters. These systems employ typically a thermal neutron sensor embedded in a High Density PolyEthylene (HDPE) moderator. Commercial devices typically utilize proportional neutron counters, gas ionization detectors employing ${}^3$He or BF${}_3$ as the active medium, and operating in the proportional region. The moderator geometry is designed to replicate the energy-dependent shape of the recommended conversion factor function between neutron fluence and ambient dose equivalent ($H^{*}(10)$). Ideally, in an ambient dosimeter, the recorded dose is  proportional to the number of events detected by the counter. However, achieving a perfect match in response shape between a moderated proportional counter and the conversion factors is unattainable in practice. Commercial dosimeters tend to overestimate the dose in the intermediate energy region (0.1 eV - 100 keV) and significantly underestimate doses for energies exceeding 10 MeV or 20 MeV, depending on their design. Certification standards account for these potential deviations by specifying tolerance ranges in the energy response of such dosimeters \cite{IEC61005}. To address energies greater than 20 MeV, extended energy range dosimeters employ high atomic number neutron multiplying materials, thereby enhancing dosimetric sensitivity for energies larger than 20 MeV \cite{birattarietal1998_extended}.

Most of the commercial non-extended and extended energy range dosimeters have been designed or adapted in the late 1990s or early 2000s. The response of these dosimeters has been designed according to the specification of ICRP 74 \cite{ICRP74}. These devices use an analog electronic chain consisting of a power module, preamplifier, signal amplifier or shaper, ADC (optional), and a counting rate monitor. The readout unit consists of an analog or digital display that converts the neutron counting rate into ambient dose equivalent rate by means of a linear calibration factor. In table \ref{tab:comp_dosimeters} is provided a compilation of the main commercial dosimeters routinely used in industrial facilities, such as nuclear reactors, processing or storage plants for combustible materials, irradiation laboratories, as well as in research facilities, such as particle accelerators, or medical facilities. Despite compilation in table \ref{tab:comp_dosimeters} is not exhaustive, it provides a summary of the state-of-the-art technology in ambient neutron dosimeters.

\begin{table}
    \centering
    \resizebox{\textwidth}{!}{
    \begin{tabular}{lccccc}
    \hline
        Manufacturer & Detector  & Release  & Mass/kg & Energy range & Upper limit \\
                     &      name  & year   &      &             &  linear response  \\  
          &         &    &      &             & in pulsed fields\\  
          &         &    &      &             & nSv/burst\\  
          \hline
        KWD Nuclear  & 2222A He-3  & Early 80's  & 10.5 & Thermal - 17 MeV & 10\\
        Instruments  &         & Updated in 2000 &      &             & \\  
        Fluke biomedical & Victoreen  & Late 90's & 9.5 & Thermal - 20 MeV & --\\
                                         &   Model 190N      &    &      &             & \\  
        Thermo Scientific & WENDI-II & 2000 & 14 & Thermal - 5 GeV & 10\\
        Berthold Technologies & LB 6411 & 1997 & 10.5 &  Thermal - 20 MeV & 10\\
        NE Technology & NM2B - 495Pb & 2002 & 18 & Thermal - 10 GeV & --\\
        ELSE nuclear & LUPIN BF3 & 2014 & 18 & Thermal - 5 GeV & 500\\
    \hline
    \end{tabular}}
    \caption{Summary compilation of the main commercially available ambient neutron dosimeters.}
    \label{tab:comp_dosimeters}
\end{table}

In the last years, some concerns have arisen regarding the reliability of portable ambient neutron dosimeters in modern facilities, especially those producing neutron fields with high-energy contributions ($E >$ 20 MeV) and complex time structures \cite{farahetal2015_measurement}. This issue is significant in medical facilities for hadron therapy with secondary stray radiation up to hundred MeVs, synchrotron or cyclotron facilities with pulsed beams or beam losses, and pulsed facilities for research and applications (e.g., spallation neutron sources, fusion experiments, high-intensity pulsed lasers). In the case of pulsed neutron fields, an EURADOS intercomparison exercise has concluded that most of the commercially available active neutron dosimeters do not perform well for pulsed dose rates higher than 10 nSv/burst \cite{caresanaetal2014_intercomparison} (see table \ref{tab:comp_dosimeters}).  Furthermore, portability is also a significant limitation of commercial neutron dosimeters. The high mass of commercial devices ($>$ 9 kg, see table  \ref{tab:comp_dosimeters}) imposes restrictions on their transportation and operation. This can pose an occupational risk for radiation protection officers who may need to handle these devices manually.

\section{The LINrem project}
\begin{figure}[t]
\centering
\includegraphics[width=\textwidth]{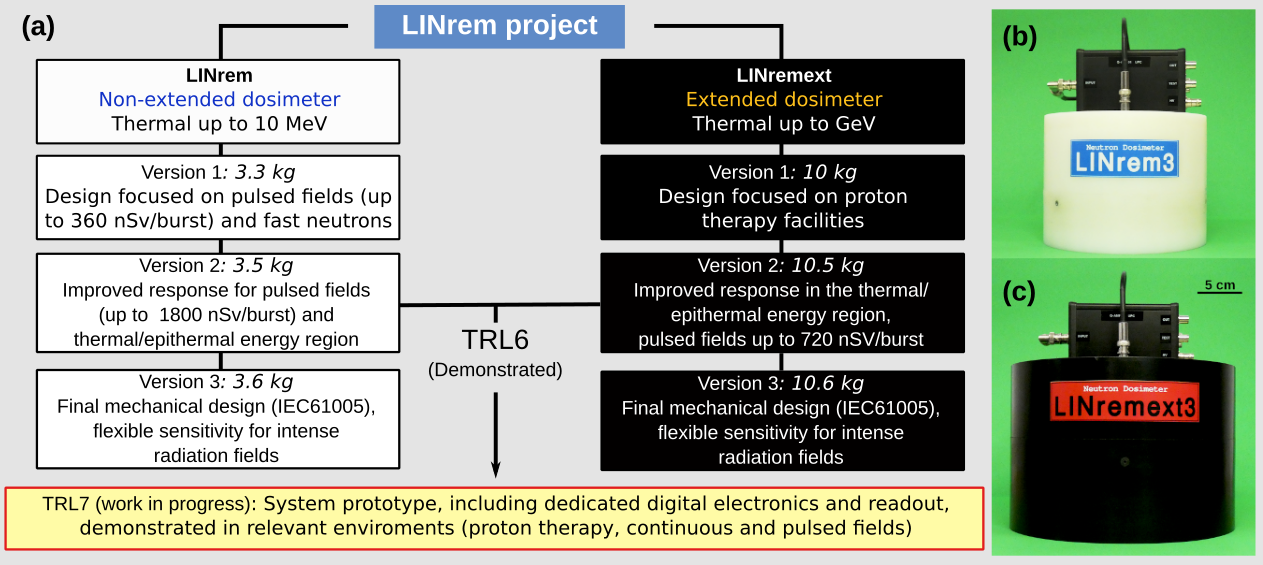}
\caption{Status of the LINrem project: (a) flowchart of the prototype versioning and TRL status. In (b) and (c) are shown pictures of the prototypes LINrem3 and LINremext3, respectively.}\label{LINrem_flowchart}
\end{figure}

The LINrem project aims to develop modern and reliable neutron radiation sensors addressing the limitations of commercial solutions for ambient neutron dosimetry in high energy and pulsed neutron fields. The key aspects in the technical proposal of the LINrem project are:
\begin{itemize}
    \item \textbf{Design:} Numerically assisted optimization of the detector design was conducted using state-of-the-art Monte Carlo (MC) transport codes. This approach enables enhanced portability and energy response of the detector.
    \item \textbf{Digital electronics:} The power, signal shaping, and readout functionalities are implemented using digital electronics, ensuring compatibility with apps and external databases. This approach offers flexibility for seamless integration with other radiation monitoring systems.
    \item \textbf{Digital data acquisition:} The acquisition process relies on digital signal processing, enabling list-mode data collection. This approach allows for adaptive processing of raw data, tailored for complex radiation fields, whether quasi-continuous or pulsed neutron fields.
    
\end{itemize}
The status of the LINrem project is depicted in a flowchart in Fig. \ref{LINrem_flowchart}. There are two design options addressing different technical specifications. The \textit{LINrem} design corresponds to a lightweight neutron dosimeter with energy sensitivity from thermal neutrons up to 10 MeV. This detector design provides a very portable solution (total mass $<$ 4 kg) for radiation protection in industry and research labs, such as nuclear power plants, irradiation facilities, technologies for non-destructive assays, low energy particle accelerators for basic research and applications, among others. The \textit{LINremext} design is an extended energy range dosimeter optimized for application in hadron therapy and Big Science projects. It provides  a sensitivity from thermal neutrons up to 10 GeV. This design also offers an improved portability (total mass $<$ 11 kg) in comparison with commercial extended dosimeters. Demonstrator prototypes for LINrem/LINremext dosimeters have been already developed and tested in relevant environments using modular electronics. Currently, a system prototype integrating the detector module, electronics and data acquisition is under development for demonstration of \textit{Technology Readiness Level} 7 (TRL7).
  
\section{Overview of the development of the neutron dosimeters and expected performance}
The development of both neutron dosimeters was started by considering a cylindrical thermal neutron proportional counter as the active sensor for the dosimeter. In particular, it was considered a proportional counter based on a ${}^3$He-filled tube, 1 inch external diameter, 5 cm active length and 10 atm internal gas pressure. This counter is manufactured and distributed by \textit{LND inc.} under model number LND-252250 \cite{LND252250}. A neutron moderator made of HDPE and cylindrical geometry is chosen by symmetry arguments.

A key aspect for the development was related to the directional effects in the detector response. For an omnidirectional radiation field, there is a compromise between the
directional effects, the moderator geometry (radius and length) and the total mass. This compromise has been studied by MC simulations using the Geant4 application \textit{ParticleCounter} \cite{ParticleCounter} yielding a parametric description of the cylinder aspect ratio while minimizing the total detector mass and constraining possible deviation of the detector response by directional effects. In a second step of the development, the optimization of neutron filters inside the non-extended design was carried out by MC simulations. For the extended version, the optimization was done by inserting neutron filters and a layer of neutron multiplier material in order to increase the sensitivity to neutron energies higher than 10 MeV. The final design of both dosimeter types required up to three iterations in prototyping (see flowchart in fig. \ref{LINrem_flowchart}). Each iteration was driven by technical validations in operational workplaces addressing improvements in the energy response and pulsed field capabilities. A summary of the nominal specifications for both dosimeter types at version 2 are presented in table \ref{tab:Sum_DosimetersV2}.

\begin{table}
    \centering
     \resizebox{\textwidth}{!}{
    \begin{tabular}{l|cc}
        \hline
        Dosimeter name & LINrem  & LINremext \\
        \hline
        Design version & 2 & 2\\
        Energy sensitivity & Thermal - 10 MeV & Thermal - 10 GeV\\
        Response variation by directional& up to 20\% & up to 20\%\\
        effects in workplaces         &           &               \\
         Systematic uncertainty by  & up to 20\% (fast) & up to 20\% (fast)\\
         neutron energy response& up to 60\% (thermal) & up to 60\% (thermal)\\
         Total dosimeter mass & $ < 4$ kg &  $ < 11$ kg \\
         Upper limit linear response & 1800 & 720\\
         in pulsed fields (nSv/burst) & & \\
         \hline
    \end{tabular}
    }
    \caption{Nominal specifications for neutron dosimeters LINrem and LINremext version 2.}
    \label{tab:Sum_DosimetersV2}
\end{table}

\section{Dosimetry in pulsed neutron fields}
The challenges associated with pulsed neutron fields arise from the rapid detection of multiple events in the detector, leading to event pile-up. Conventional nuclear instrumentation struggles to separate simultaneous events in time (pulse mode) or to register a steady current proportional to the counting rate (current mode) under these conditions, resulting in significant underestimations. Given the instantaneous nature of pulsed fields, the key metric for event counting in this scenario is the integrated number of events detected per radiation burst.

In conditions of intense pile-up in proportional neutron counters, assuming no variations in gas gain due to spatial charge accumulation, the total charge collected after a radiation burst is proportional to the total number of events detected per burst. This operation mode, known as ``charge integration", combines aspects of both current and pulse modes. Originally proposed in 2008 for monitoring neutron yield in transient fusion plasmas \cite{morenoetal2008}, the charge integration mode was implemented using charge-sensitive preamplifiers. Subsequently, this methodology was mathematically formalized to estimate counting uncertainties associated with calibration and use of proportional counters in pulsed neutron fields \cite{tarifeno-saldiviaetal2010_1,tarifeno-saldiviaetal2014_calibration}. The formalization involved developing a statistical model that comprehensively considers the pile-up process, the conversion of collected charge into detected events, and the uncertainties associated with event counting in the charge integration mode.

The LUPIN-BF3 dosimeter (see Table \ref{tab:comp_dosimeters}) employs current-sensitive preamplifiers to operate in the charge integration mode. During its prototype phase in 2013, this detector exhibited non-linear response deviations for pulsed doses exceeding 16 nSv/burst \cite{caresanaetal13_lupin}, attributed to gain losses from spatial charge accumulation within the counter. A study conducted in 2013 on the effects of spatial charge accumulation in proportional counters, used for neutron yield monitoring in plasma sources, yielded an effective correction model for these gain losses \cite{riosetal2013_total}. Implementing this correction model into the processing firmware of LUPIN-BF3 extended its linearity range beyond 100 nSv/bunch \cite{casselletal2015_novel}. Presently, LUPIN-BF3 stands as the sole commercially available active dosimeter with capabilities for pulsed fields exceeding 10 nSv/bunch  \cite{caresanaetal2014_intercomparison}.

\begin{figure}[t]
\centering
\includegraphics[width=0.85\textwidth]{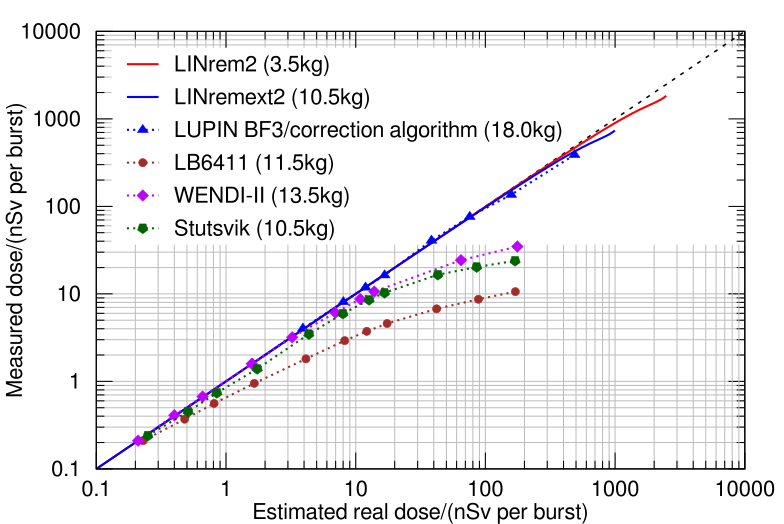}
\caption{Performance of LINrem2 and LINremext2 dosimeters in pulsed neutron fields. Data from dosimeters LUPIN BF3, LB6411, WENDI-II and Stutsvik has been extracted from \cite{caresanaetal2014_intercomparison}.}
\label{Comparison_pulsedFields}
\end{figure}

The LINrem and LINremext dosimeters implement charge integration mode using charge-sensitive preamplifiers. An overview of the performance of both dosimeters in pulsed neutron fields is presented in Fig. \ref{Comparison_pulsedFields} and compared to commercial devices as reported in an intercomparison exercise conducted by EURADOS \cite{caresanaetal2014_intercomparison}. Remarkably, LINrem and LINremext dosimeters demonstrate reliable performance in pulsed neutron fields, exhibiting a linear response beyond the upper limit of LUPIN-BF3, and offering unsurpassed portability features. The lightweight non-extended energy range version weighs just 3.5 kg, while the competitive extended version weighs 10.5 kg. The application of both dosimeters for radioprotection in pulsed neutron fields is discussed in the following sections.

\subsection{Pulsed neutron fields from fusion plasmas}

\begin{figure}[t]
\centering
\includegraphics[width=\textwidth]{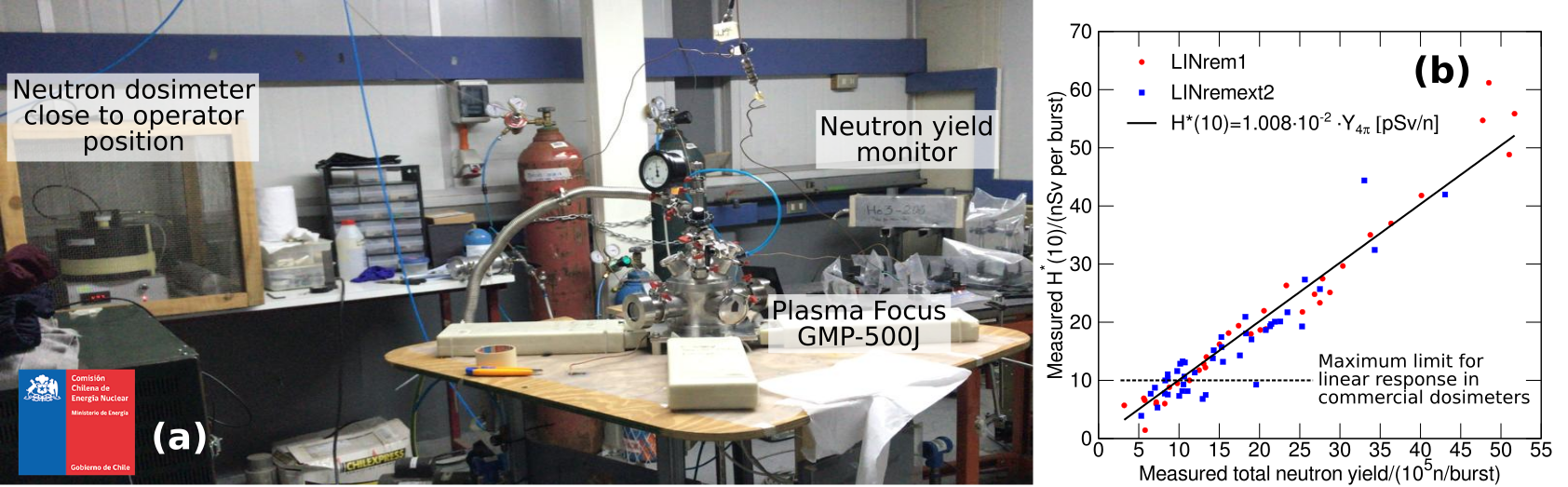}
\caption{Dosimetric characterization of a pulsed neutron field workplace at the Chilean Nuclear Energy Commission. The experimental setup is depicted in panel (a). In panel (b), the correlation between measurements of $H^{*}(10)$ and the total neutron yield is presented.}
\label{Exp_CCHEN}
\end{figure}

The dosimetric characterization of a pulsed neutron field based on fusion plasmas has been achieved in collaboration with the Chilean Nuclear Energy Commission (CCHEN). At CCHEN, the research program is focused on the development of low-energy Plasma Focus (PF) discharges for fundamental physics studies related to nuclear fusion and for the production of pulsed fields of X-rays and neutron radiation. These efforts target materials studies, radiobiology, and microdosimetry applications.

In a PF discharge, operated with deuterium gas at mbar pressures, D-D fusion reactions occur via magnetic confinement of a plasma. Plasma confinement times are typically tens of nanoseconds, resulting in bursts of X-ray radiation (from electron bremsstrahlung) and fusion neutrons ($\sim$ 2.5 MeV). The PF neutron source is driven by the Multipurpose Generator (MPG) \cite{tarifenoetal2008_design}, a tabletop pulsed power generator (100 kA, 290-345 J) capable to drive a PF producing $10^5$-$10^7$  n/burst in single shot regime \cite{pavezetal2023_new}.

Given that the tabletop PF operates within a small lab room (around $4\times 3$  $m^2$) and that the operator may need to be inside the room during experimental activities, a key aspect of this work was to determine the pulsed dose rate at the operator's position inside the room. 

The experimental setup is depicted in Fig. \ref{Exp_CCHEN}, alongside the experimental results. The experimental campaign involved conducting independent ambient dose measurements using LINrem1 and LINremext2, which were then correlated with separate measurements of the total neutron yield for the same shot (see Fig. \ref{Exp_CCHEN}.b). The outcome of the experiment indicate that the pulsed dose rate exhibits a linear relationship with the total neutron yield, surpassing the limit of 10 nSv/burst for non-linear response in most commercial ambient neutron dosimeters. Based on the experimental results, the recommended conversion factor from total neutron yield to ambient dose equivalent at the operator's position is $H^{*}(10)=10^{-2}\cdot Y_{4\pi}$ pSv/neutron/burst.

Given that the device typically produces a few hundred shots per day, and assuming an average neutron yield of $10^6 neutrons/burst$, it is concluded that on a normal workday, the operator's total exposure will be lower than 8 $\mu Sv/day$, which is acceptable for occupational exposure.

\subsection{Pulsed neutron fields from laser driven neutron sources}
The rapid development of ultra-short (femtosecond) and high-power lasers now enables the acceleration of ions, which can subsequently generate pulsed neutron fields. The neutron yield achieved are comparable to, and in some cases even higher than, those obtained with conventional accelerators, making them highly appealing to the neutron beam user community.

The VEGA-3 laser system at the Centro de Láseres Pulsados (CLPU), in Spain, delivers laser pulses up to 1 PW at a repetition rate of 1 Hz, up to 30 J per shot, 30 fs laser pulse duration and focused to intensities up to $2.5 \times 10^{20} W cm^{-2}$. The assessment of pulsed neutron ambient dose equivalent in this facility was conducted in collaboration with researchers from CLPU and the Spanish nuclear physics community. In particular, the LINrem2 and LINremext2 detectors were used during the commissioning of the CLPU neutron source, based on laser-driven proton acceleration into a ${}^7$Li  target with a total laser pulse energy of 25 J per shot. This neutron source is expected to produce around $10^8$ neutrons per burst with energies ranging from 100 keV up to 10 MeV. The experimental setup is depicted in Fig. 4, where the neutron dosimeters were positioned at an angle of 40 degrees with respect to the laser beam axis and approximately 2.5 m from the production target. Experimental results for time-series data from two different days during the experimental campaign are presented in Fig. 5. 
For these measurements, the net neutron dose was obtained after subtracting the contribution of the gamma-flash and electromagnetic pulse into the detector. This contribution was estimated from shots with a Polyethylene target, where negligible neutron production is expected, and reached around 50\% for LINrem2 and 20\% for LINremext2, respectively. At the measurement position, the pulsed dose rates typically range from 10 to 30 nSv per burst. Therefore, based on our measurements, the recommended value is,
\begin{equation*}
    \overline{H^{*}}(10) = 20\quad nSv/burst,
\end{equation*}
serving as a reference for comparison with Monte Carlo simulations of the facility and the assessment of exposure risk due to neutron radiation inside the bunker.

\begin{figure}[t]
\centering
\includegraphics[width=\textwidth]{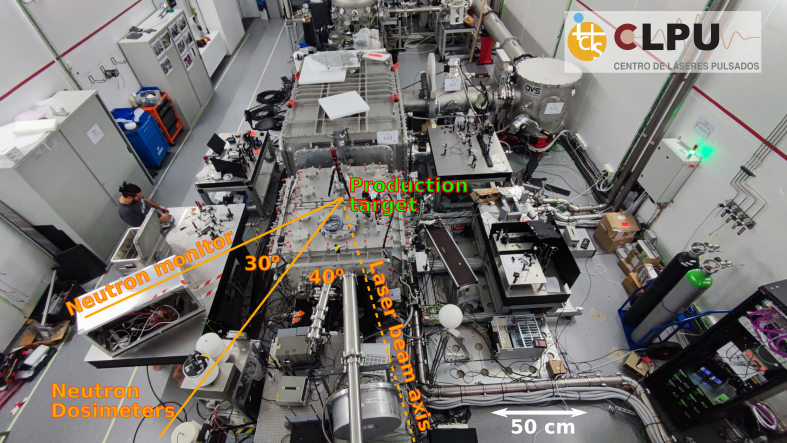}
\caption{Experimental setup during commissioning of the laser-driven neutron source at CLPU.}
\label{Scheme_CLPU}
\end{figure}

\begin{figure}[t]
\centering
\includegraphics[width=\textwidth]{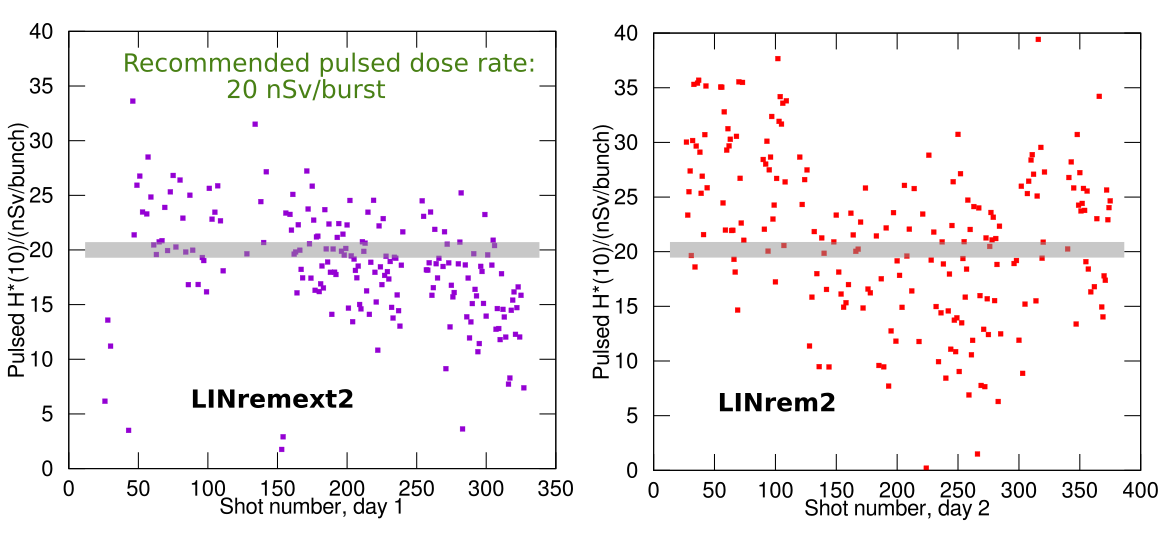}
\caption{Pulsed neutron ambient equivalent rate at CLPU using LINremext2 and LINrem2. Data series from two different days during commissioning of the laser-driven neutron source.}
\label{Results_doses_CLPU}
\end{figure}


\section{Final remarks and future perspectives}
The status of the LINrem project  has been outlined, detailing the development of two new dosimeter types: LINrem and LINremext. LINrem offers a lightweight, portable solution for radiation protection across various facilities, spanning nuclear power plants, research labs, and particle accelerators. Covering an energy range from thermal neutrons to 10 MeV, LINrem weighs under 4 kg. LINremext, tailored for hadron therapy and Big Science projects, extends the energy range to 10 GeV with a total mass under 11 kg, suitable for diverse applications. These dosimeters have been successfully applied for radioprotection in pulsed neutron fields from fusion plasmas and laser-driven sources. Ongoing work includes their application in high-energy proton therapy facilities, with results to be presented separately. Additionally, exploration of their potential in medical facilities for ultra-high dose rate therapy is underway. As for future perspectives, efforts are underway to protect intellectual property for unique detector design aspects \cite{linremPatent}. Development is ongoing for a system prototype integrating compact electronics and fully digital readout. CE marking is the next milestone, aimed at bringing the development closer to end-users.

\section{Acknowledgements}
This work has been supported by Ag\`encia de Gesti\'o d’Ajuts Universitaris i de Recerca (AGAUR) grant 2018-LLAV-00039. Grant ACM-2019-014 Universidad Polit\'ecnica de Catalunya-FEDER. Grants from Ministerio de Ciencia e Innovacion  PID2019-104714GB-C21, PID2019-104714GB-C22 and PDC2021-121536-C21/2. F. M. acknowledges support from ANID FONDECYT Regular Project 1221364. C. P. acknowledges support from ANID-Fondecyt project N°1211885. The authors would like to express their gratitude to the technical and scientific staff at CLPU for their support during the realization of the experimental activities there.



\bibliographystyle{elsarticle} 





\end{document}